\documentstyle[thmsa,11pt,a4,sw20lart]{article}

\input tcilatex
\begin{document}

\topmargin 0pt \oddsidemargin 5mm

\setcounter{page}{1}

\hspace{8cm}{} \vspace{2cm}

\begin{center}
{\large {STOPPING OF CHARGED PARTICLES IN A MAGNETIZED CLASSICAL PLASMA}}\\

{H.B. Nersisyan}\\\vspace{1cm} {\em Division of Theoretical Physics,
Institute of Radiophysics and Electronics, 2 Alikhanian Brothers St.,
Ashtarak-2, 378410, Republic of Armenia}\footnote{%
E-mail: Hrachya@irphe.am}\\
\end{center}

\vspace {5mm} \centerline{{\bf{Abstract}}}

The analytical and numerical investigations of the energy loss rate of the
test particle in a magnetized electron plasma are developed on the basis of
the Vlasov-Poisson equations, and the main results are presented. The Larmor
rotation of a test particle in a magnetic field is taken into account. The
analysis is based on the assumption that the energy variation of the test
particle is much less than its kinetic energy. The obtained general
expression for stopping power is analyzed for three cases: (i) the particle
moves through a collisionless plasma in a strong homogeneous magnetic field;
(ii) the fast particle moves through a magnetized collisionless plasma along
the magnetic field; and (iii) the particle moves through a magnetized
collisional plasma across a magnetic field. Calculations are carried out for
the arbitrary test particle velocities in the first case, and for fast
particles in the second and third cases. It is shown that the rate at which
a fast test particle loses energy while moving across a magnetic field may
be much higher than the loss in the case of motion through plasma without
magnetic field.

\newpage 

\section{INTRODUCTION}

Energy loss of fast charged particles in a plasma has been a topic of great
interest since the 1950s [1-8] due to its considerable importance for the
study of basic interactions of the charged particles in real media;
moreover, recently it has also become a great concern in connection with
heavy-ion driven inertial fusion research [5-8].

The nature of experimental plasma physics is such that experiments are
usually performed in the presence of magnetic fields, and consequently it is
of interest to investigate the effects of a magnetic field on the energy
loss rate. Strong magnetic fields used in the laboratory investigations of
plasmas can appreciably influence the processes determined by Coulomb
collisions [9]. This influence is even more important in white dwarfs and in
neutron stars, the magnetic fields on the surfaces of which can be as high
as $10^5-10^{10}\ {\rm kG}$.

Stopping of charged particles in a magnetized plasma has been the subject of
several papers [10-15]. Stopping of a fast test particle moving with
velocity $u$ much higher than the electron thermal velocity $v_T$ was
studied in Refs. [10,11,13]. Energy loss of a charged particle moving with
arbitrary velocity was studied in Ref. [12]. The expression obtained there
for the Coulomb logarithm, $L=\ln (\lambda _D/\rho _{\perp })$ (where $%
\lambda _D$ is the Debye length and $\rho _{\perp }$ is the impact parameter
for scattering for an angle $\vartheta =\pi /2$), corresponds to the
classical description of collisions. In the quantum-mechanical case, the
Coulomb logarithm is $L=\ln (\lambda _D/\lambda _B)$, where $\lambda _B$ is
the de Broglie wavelength of plasma electrons [16].

In Ref. [15], the expressions were derived describing the stopping power of
a charged particle in Maxwellian plasma placed in a classically strong (but
not quantizing) magnetic field ($\lambda _B\ll a_c\ll \lambda _D,$ where $%
a_c $ is the electron Larmor radius), under the conditions when scattering
must be described quantum mechanically. Calculations were carried out for
slow test particles whose velocities satisfy the conditions $%
(m_e/m_i)^{1/3}v_T<u\ll v_T$, where $m_i$ is the mass of the plasma ions and 
$m_e$ is the electron mass.

The reaction of a uniform plasma to an electrostatic field of a moving test
particle was studied by Rostoker and Rosenbluth [17] for two cases, in the
presence or absence of a uniform magnetic field. For a test particle having
velocity $u\gg v_T$ in the positive $z$ direction, the dielectric function
for no magnetic field gives a resonance $k_z=\omega _p/u$. As a result, the
emission of plasma waves by the test particle with given $k$ is concentrated
on the cone forming an angle $\theta $ with respect to $u$, where $\cos
\theta =k_z/k$. As shown by Rostoker and Rosenbluth, this leads to a $%
\stackrel{\vee }{\text{C}}$herenkov-type shock front making the acute angle $%
\pi /2-\theta $ with the negative $z$ axis. Their treatment in the presence
of a magnetic field was very general and involved no assumption concerning
the relative magnitudes of $\omega _p$ and $\omega _c$, i.e., the electron's
plasma and cyclotron frequencies. Stopping power was not determined for any
specific case. The authors were aware that in the case when $\omega _c\gg
\omega _p$, where field electrons in the lowest order can respond only to
the waves in the direction of ${\bf B}_0$, the resonance caused by the
dielectric function has a different form, with $k=\omega _p/u$, being
independent of the ${\bf k}$ direction. The electrostatic field of a moving
test particle in such magnetized plasma was studied by Ware and Wiley [18].

In the present paper we calculate, in a framework of dielectric theory, the
energy loss rate of a test particle moving in magnetized plasma. We consider
the test particle interaction only with the electron component of plasma,
since it is this interaction that dominates the stopping of a test particle
[19,20]. Besides, in contrast with the papers [10-15], Larmor rotation of a
test particle in a magnetic field is taken into account.

In Sec. II, linearized Vlasov-Poisson equations are solved by means of
Fourier analysis in order to obtain a general form for the linearized
potential generated in a magnetized Maxwellian plasma by a test particle and
for the energy loss rate of the test particle.

In Sec. III, the energy loss in a Maxwellian collisionless plasma in the
presence of a strong magnetic field is examined. Calculations are carried
out for arbitrary test particle velocities. In this case, plasma
oscillations are also excited, though their spectra in the strong magnetic
field differ from normal ones.

In Sec. IV, the energy loss rate in a cold plasma is calculated in the case
when the fast particle moves along ($\vartheta =0$) and across ($\vartheta
=\pi /2$) the magnetic field. It is shown that in the first case, the energy
loss rate is less than Bohr's result. In the second case, the energy loss
rate can be much higher than Bohr's result.

In Sec. V, we present a qualitative discussion of obtained results. In the
Appendix, analysis of the function $Q_\nu (z)$ is given.

\section{BASIC RELATIONS}

We consider a nonrelativistic charged particle having charge $Ze$ that moves
in a magnetized plasma at an angle $\vartheta $ with respect to the magnetic
field directed along the $z$ axis. We assume that the energy variation of
the particle is much smaller than its kinetic energy. In this case the
charge density associated with the test particle is given by the following
expression:

\begin{equation}
\rho _0({\bf r},t)=Ze\delta (x-a\sin (\Omega _ct))\delta (y-a\cos (\Omega
_ct))\delta (z-u_0t),
\end{equation}
where $u_0$ and $v$ are the particle velocity components along and across
from the magnetic field ${\bf B}_0$ ($u_0=u\cos \vartheta ,$ $v=u\sin
\vartheta $), where $u$ is the particle velocity, $\Omega _c=ZeB_0/Mc$, $%
a=v/\Omega _c$, and $M$ are the Larmor frequency, the Larmor radius, and the
mass of the particle, respectively, and $c$ is the speed of light.

The linearized Vlasov equation of the plasma may be written as

\begin{equation}
\frac{\partial f_1}{\partial t}+{\bf v}\frac{\partial f_1}{\partial {\bf r}}%
+\omega _c\left[ {\bf v\times b}_0\right] \frac{\partial f_1}{\partial {\bf v%
}}=\frac em\frac{\partial \varphi }{\partial {\bf r}}\frac{\partial f_0}{%
\partial {\bf v}},
\end{equation}
where the self-consistent electrostatic potential $\varphi $ is determined
by Poisson's equation

\begin{equation}
\nabla ^2\varphi =-4\pi \rho _0({\bf r},t)-4\pi e\int d{\bf v}f_1({\bf r},%
{\bf v},t),
\end{equation}
where ${\bf b}_0$ is the unit vector parallel to ${\bf B}_0$, $e$, $m$, and $%
\omega _c$ are the charge, mass, and Larmor frequency of plasma electrons,
respectively, $f_0$ is the unperturbed distribution function of plasma
electrons, which is taken uniform and Maxwellian,

\begin{equation}
f_0(v)=\frac{n_0}{\left( 2\pi v_T^2\right) ^{3/2}}\exp \left( -\frac{v^2}{%
2v_T^2}\right) ,
\end{equation}
with $v_T=\sqrt{k_BT/m}$. Here, $n_0$ is the unperturbed number density of
the plasma electrons.

By solving Eqs. (2) and (3) in space-time Fourier components, we obtain the
following expression for the electrostatic potential:

\begin{eqnarray}
\varphi ({\bf r},t) &=&\frac{Ze}\pi \sum_{n=-\infty }^\infty \exp [-in(\psi
+\Omega _ct)]\int_0^\infty dk_{\perp }k_{\perp }J_n(k_{\perp }a)J_n(k_{\perp
}\rho ) \\
&&\times \int_{-\infty }^{+\infty }\frac{dk_z\exp (ik_z\xi )}{\left(
k_z^2+k_{\perp }^2\right) \varepsilon (k_z,k_{\perp },k_zu_0+n\Omega _c)}, 
\nonumber
\end{eqnarray}
where $k^2=k_z^2+k_{\perp }^2$, $\xi =z-u_0t$, $J_n$ is the $n$th order
Bessel function, $\rho $, $\psi $, and $z$ are the cylindrical coordinates
of the observation point, and $\varepsilon (k_z,k_{\perp },\omega )$ is the
plasma dielectric function, which has been given by many authors [21,22],
and may be written in the form

\begin{equation}
\varepsilon (k_z,k_{\perp },\omega )=1+\frac 1{k^2\lambda _D^2}\left[
1+2ip\int_0^\infty dt\exp \left( 2ipt-W\right) \right]
\end{equation}
with $p=\omega /\sqrt{2}kv_T$ and

\begin{equation}
W=t^2\cos ^2\alpha +k^2a_c^2\sin ^2\alpha \left[ 1-\cos \left( \frac{\sqrt{2}%
\omega _ct}{kv_T}\right) \right] .
\end{equation}
Here, $\alpha $ is the angle between the wave vector ${\bf k}$ and the
magnetic field.

The result represents a dynamical response of the medium to the motion of
the test particle in the presence of the external magnetic field; it takes
the form of an expansion over all the harmonics of the Larmor frequency of
the particle.

The energy loss rate (ELR) $S$ of a fast charge is defined as the energy
loss of the charge in a unit time due to interactions with the plasma
electrons. From Eq. (5) it is straightforward to calculate the electric
field ${\bf E}({\bf r},t)=-{\bf \nabla }\varphi ({\bf r},t)$, and the
stopping force acting on the particle. Then, the ELR of the test particle
becomes

\begin{eqnarray}
S &=&\frac{2Z^2e^2}\pi \sum_{n=-\infty }^\infty \int_0^{k_{\max }}dk_{\perp
}k_{\perp }J_n^2(k_{\perp }a)  \nonumber \\
&&\times \int_0^\infty dk_z\frac{k_zu_0+n\Omega _c}{k_z^2+k_{\perp }^2}{\rm %
Im}\frac{-1}{\varepsilon (k_z,k_{\perp },k_zu_0+n\Omega _c)},
\end{eqnarray}
with $k_{\max }=1/r_{\min }$, where $r_{\min }$ is the effective minimum
impact parameter. Here $k_{\max }$ has been introduced to avoid the
divergence of the integrals caused by the incorrect treatment of the
short-range interaction between the test particle and the plasma electrons
within the linearized Vlasov theory.

\section{ELR IN PLASMA IN THE PRESENCE OF A STRONG MAGNETIC FIELD}

Let us analyze expression (8) in the case when a particle moves in a plasma
with a sufficiently strong magnetic field. Let us assume the magnetic field,
on one hand, reasonably weak and not to be quantized ($\hbar \omega _c<k_BT$
or $a_c\gg \lambda _B$), and, on the other hand, comparatively strong so
that the cyclotron frequency of the plasma electrons exceeds the plasma
frequency $\omega _p=\sqrt{4\pi n_0e^2/m}$ or $a_c\ll \lambda _D$, where $%
a_c $ is the Larmor radius and $\lambda _D$ is the Debye length $\lambda
_D=v_T/\omega _p$. Because of this assumption, the perpendicular cyclotron
motion of the test and plasma particles is neglected. The test particle's
velocity parallel to ${\bf B}_0$ is taken as $u_0$. The generation of an
electrostatic wake by a superthermal test electron in a magnetized electron
plasma in this limit has been discussed by Ware and Wiley [18].

In the limit of sufficiently strong magnetic field, Eq. (8) becomes

\begin{equation}
S=\frac{2Z^2e^2u_0}\pi \int_0^{k_{\max }}dk_{\perp }k_{\perp }\int_0^\infty
dk_z\frac{k_z}{k^2}{\rm Im}\frac{-1}{\varepsilon _\infty (k_z,k_{\perp
},k_zu_0)}
\end{equation}
with

\begin{equation}
\varepsilon _\infty (k_z,k_{\perp },\omega )=1+\frac 1{k^2\lambda
_D^2}\left[ X\left( \frac \omega {k_zv_T}\right) +i\frac{\left| k_z\right| }{%
k_z}Y\left( \frac \omega {k_zv_T}\right) \right] ,
\end{equation}
where $W(z)=X(z)+iY(z)$ is the plasma dispersion function [23],

\begin{equation}
X(z)=1-\sqrt{2}zDi\left( \frac z{\sqrt{2}}\right) ,
\end{equation}
\begin{equation}
Y(z)=\sqrt{\frac \pi 2}z\exp \left( -\frac{z^2}2\right) ,
\end{equation}
\begin{equation}
Di(z)=\exp \left( -z^2\right) \int_0^zdt\exp \left( t^2\right)
\end{equation}
is the Dawson integral [23]. At large values of its argument, the Dawson
integral has the value $Di(z)\simeq 1/2z+1/4z^3$.

Substituting Eq. (10) into Eq. (9) and making the substitutions $\lambda
=u_0/v_T,$ $B=k_{\max }\lambda _D$ we obtain

\begin{eqnarray}
S &=&\frac{Z^2e^2v_T}{2\pi \lambda _D^2}\lambda \left\{ \frac{Y(\lambda )}%
2\ln \frac{Y^2(\lambda )+\left( B^2+X(\lambda )\right) ^2}{Y^2(\lambda
)+X^2(\lambda )}\right. \\
&&\ +B^2\left[ \frac \pi 2-\arctan \frac{B^2+X(\lambda )}{Y(\lambda )}\right]
\nonumber \\
&&\ \left. +X(\lambda )\left[ \arctan \frac{X(\lambda )}{Y(\lambda )}%
-\arctan \frac{B^2+X(\lambda )}{Y(\lambda )}\right] \right\} .  \nonumber
\end{eqnarray}

The maximum value of $k_{\perp }$, $k_{\max }$, will be $a_c^{-1}$ for
fusion plasmas, since the magnetized plasma approximation that neglects the
perpendicular motion of electrons ceases to be valid for collision
parameters less than $a_c$.

The first term of Eq. (14) is a contribution to the frictional drag due to
collisions with the plasma electrons. It is incomplete because the analysis
treats the background electrons as a continuous fluid and there is no
allowance being made for the recoil of the test particle due to each
collision. The other terms are associated with the resonance giving rise to
plasma wave emission.

From Eqs. (9)-(14) we can assume that the main contribution in the ELR is
given by the values of the particle's velocity, for which $X(\lambda )<0$
and $Y(\lambda )\ll \left| X(\lambda )\right| $. These conditions correspond
to excitation of plasma waves by a moving particle. As shown by Rostoker and
Rosenbluth [17], the plasma waves were not determined for any specific case.
They were aware that for the case $\omega _c\gg \omega _p$, where the plasma
electrons in the lowest order can respond to the waves only in the direction
of ${\bf B}_0$, the resonance caused by the dielectric function has a
different form $k=\omega _p/u_0$, being independent of the ${\bf k}$
direction. Plasma waves involved in this case are oblique plasma waves
having the approximate dispersion relation $\omega _k=\omega _pk_z/k$. In
Secs. III A and III B the expression (14) is evaluated for large and small
test particle velocities.

\subsection{ELR for small velocities}

When a test particle moves slowly through a plasma, the electrons have much
time to experience the particle's attractive potential. They are accelerated
towards the particle, but when they reach its trajectory the particle has
already moved forward a little bit. Hence, we expect an increased density of
electrons at some place in the trail of the particle. This negative charge
density pulls back the positive particle and gives rise to the ELR.

The Taylor expansion of Eq. (14) for small $u_0$ ($\lambda \ll 1$) yields
the ``friction law''

\begin{equation}
S=\frac{Z^2e^2v_T}{2\sqrt{2\pi }\lambda _D^2}\left[ \lambda ^2{\cal R}%
_1-\lambda ^4{\cal R}_2+O(\lambda ^6)\right]
\end{equation}
with the ''friction coefficient''

\begin{equation}
{\cal R}_1=\ln \left( 1+B^2\right)
\end{equation}
and the $\lambda ^4$ coefficient

\begin{equation}
{\cal R}_2=\frac 12\ln \left( 1+B^2\right) -\frac 12\left( 1-\frac \pi
6\right) -\frac \pi 4\frac 1{\left( 1+B^2\right) ^2}+\frac \pi 6\frac
1{\left( 1+B^2\right) ^3}.
\end{equation}
Note that $B=\omega _c/\omega _p$ and therefore $B\gg 1$. The Coulomb
logarithms in Eqs. (16) and (17) are then the leading terms. We obtain

\begin{equation}
S=\frac{Z^2e^2v_T}{2\sqrt{2\pi }\lambda _D^2}\left\{ 2\lambda ^2\ln
B-\lambda ^4\left[ \ln B-\frac 12\left( 1-\frac \pi 6\right) \right]
+O(\lambda ^6)\right\} .
\end{equation}

The most important property of the ELR at small velocities is $S\propto
u_0^2 $ provided that the density is not too high ($\omega _p<\omega _c$).
This looks like the friction law of a viscous fluid, and accordingly ${\cal R%
}_1$ is called the friction coefficient. However, in the case of an ideal
plasma it should be noted that this law does not depend on the plasma
viscosity and is not a consequence of electron-electron collisions with
small impact parameter. These collisions are neglected in the Vlasov
equation. As described above, it is rather the fact that the dressing of the
test particle takes some time and produces the negative charge behind the
particle leading to the drag.

\subsection{ELR for large velocities}

For large $u_0$ ($u_0\gg v_T$) we have $X(\lambda )\simeq -1/\lambda ^2,\
Y(\lambda )\simeq 0.$ In this case Eq. (14) becomes

\begin{equation}
S\simeq \frac{Z^2e^2\omega _p^2}{2v_T}\frac{v_T}{u_0}.
\end{equation}
From Eq. (19) we can assume that the ELR is $2L=2\ln (k_{\max
}^{(0)}u_0/\omega _p)$ (where $k_{\max }^{(0)}$ is a cutoff parameter in a
plasma in the absence of magnetic field) times smaller than the Bohr ELR
[24].

Our assumption made at the beginning of this section was that the classical
approach in consideration of energy losses in plasma placed in a strong
magnetic field limits the values of the magnetic field itself and values of
temperature and plasma concentrations. From these conditions we can obtain

\begin{equation}
3\times 10^{-6}n_0^{1/2}<B_0<10^5T,
\end{equation}
where $n_0$ is measured in ${\rm cm}^{-3}$, $T$ is measured in eV, and $B_0$
in kG. Conditions (20) are always true in the range of parameters $%
n_0<10^{15}\ {\rm cm}^{-3}$, $B_0<100\ {\rm kG}$, $T>10^{-3}\ {\rm eV}$.

In Fig. 1, the ELR is plotted as a function of parameter $\lambda $ for $%
T=10\ {\rm eV}$, $n_0=10^{14}\ {\rm cm}^{-3}$, and for two different values
of $B_0$: $B_0=50\ {\rm kG}$ (dotted line) and $B_0=80\ {\rm kG}$ (solid
line). The peak corresponds to excitation of plasma waves by a moving
particle.

\section{ELR OF A FAST CHARGED PARTICLE IN\protect\\  COLD MAGNETIZED PLASMA}

We shall further analyze Eq. (8) in the case when the fast particle moves in
a cold plasma whose longitudinal dielectric function is given by the
following expressions [25,26]:

\begin{equation}
\varepsilon (k_z,k_{\perp },\omega )=\varepsilon (\omega )\cos ^2\alpha
+h(\omega )\sin ^2\alpha
\end{equation}
with

\begin{equation}
\varepsilon (\omega )=1-\frac{\omega _p^2}{\omega (\omega +i\nu )},\quad
h(\omega )=1+\frac{\omega _p^2(\omega +i\nu )}{\omega [\omega _c^2-(\omega
+i\nu )^2]},
\end{equation}
where $\nu $ is the effective collision frequency. The collisions are
negligible if the frequency of collisions with large scattering angle
between the electrons is small compared with the plasma frequency $\omega _p$%
. The cross section for collisions with scattering angles of $90^{\circ }$
or more is $\sigma _{90^{\circ }}=\pi r_{90^{\circ }}^2=\pi \left(
e^2/k_BT\right) ^2$ and the frequency of such collisions $\nu =n_0\sigma
_{90^{\circ }}v_T$. Thus

\begin{equation}
\frac \nu {\omega _p}=\frac 14\left[ \frac \pi 2n_0\left( \frac{e^2}{k_BT}%
\right) ^3\right] ^{1/2}.
\end{equation}
If $T\gg 6.6\times 10^{-8}n_0^{1/3}$, then $\nu \ll \omega _p$ and the
collisions in the plasma may be ignored.

In Eq. (8) we introduced a cutoff parameter $k_{\max }$ in order to avoid
the logarithmic divergence at large $k_{\perp }$. This divergence
corresponds to the incapability of the linearized Vlasov theory to treat
close encounters between the test particle and the plasma electrons
properly. The full nonlinear Vlasov equation accurately describes the
scattering of individual electrons with the test particle in accordance with
the Rutherford scattering theory. The exact expression for energy transfer
in the Rutherford two-body collision is

\begin{equation}
\Delta E(\rho )=\frac{(\Delta {\bf p)}^2}{2m}=\frac{2Z^2e^4}{mv_r^2}\frac
1{\left( \frac{Ze^2}{mv_r^2}\right) ^2+\rho ^2},
\end{equation}
where $v_r\simeq (u^2+v_T^2)^{1/2}$ is the mean relative velocity between
the test particle and the electron. From the denominator in Eq. (24) it
follows that the effective minimum impact parameter is $r_{\min
}=Ze^2/mv_r^2 $, which is often called the ``distance of closest approach.''
Thus,

\begin{equation}
k_{\max }=\frac 1{r_{\min }}=\frac{m\left( u^2+v_T^2\right) }{Ze^2}
\end{equation}
ensures agreement of Eq. (8) with the Rutherford theory for small impact
parameters. When $u>2Ze^2/\hbar $, the de Broglie wavelength begins to
exceed the classical distance of closest approach. Under these circumstances
we choose $k_{\max }=2mu/\hbar $.

\subsection{Longitudinal motion of a particle ($\vartheta =0$)}

In the case of an incidence angle $\vartheta =0$ of the test particle, we
obtain from Eqs. (8) and (21) the following expression:

\begin{equation}
S=\frac{2Z^2e^2}{\pi u_0}\int_0^{k_{\max }}dk_{\perp }k_{\perp
}\int_0^\infty d\omega \omega {\rm Im}\frac{-1}{k_{\perp }^2h(\omega
)+\left( \omega ^2/u_0^2\right) \varepsilon (\omega )}.
\end{equation}

Due to the resonant character of the integral over $\omega $ in the
expression (26), the main contribution to the energy losses gives those
ranges of integration where ${\rm Im}\varepsilon \ll {\rm Re}\varepsilon $
and ${\rm Im}h\ll {\rm Re}h$. These conditions are true when $\nu \ll \omega
_p$. By using the property of the Dirac $\delta $ function from expression
(26), we have

\begin{equation}
S=\frac{2Z^2e^2}{u_0}\int_0^{k_{\max }}dk_{\perp }k_{\perp }\int_0^\infty
d\omega \omega \delta \left[ k_{\perp }^2h(\omega )+\left( \omega
^2/u_0^2\right) \varepsilon (\omega )\right] .
\end{equation}

In the expression (27) the argument of the $\delta $ function defines the
frequencies of normal oscillations of a magnetized plasma in the
long-wavelength approximation. In general, they are studied in Refs. [27,28]
in more detail for electron plasma. After integration in expression (27), we
have

\begin{equation}
S=\frac{Z^2e^2}{u_0}\int_C\frac{d\omega \omega }{\left| h(\omega )\right| },
\end{equation}
where the range of integration $C$ can be determined from the inequality $%
P(\omega )<-\omega ^2/k_{\max }^2u_0^2$ and $P(\omega )=h(\omega
)/\varepsilon (\omega )$.

Integrating over frequency in the expression (28), we obtain finally

\begin{equation}
S=\frac{Z^2e^2v_T}{4\lambda \lambda _D^2}\left[ F(\beta )-\sqrt{F^2(\beta
)-4\beta ^2}+2\ln \frac{F(\beta )+\sqrt{F^2(\beta )-4\beta ^2}}{2(1+\beta ^2)%
}\right] ,
\end{equation}
where $\beta =\omega _c/\omega _p$, $\lambda =u_0/v_T$, and $F(\beta
)=1+\beta ^2+\lambda ^2B^2$, with

\begin{equation}
B=k_{\max }\lambda _D=\left\{ 
\begin{array}{c}
(k_BT/Ze^2\lambda _D^{-1})\lambda ^2,\quad 1\ll \lambda <2Ze^2/\hbar v_T, \\ 
(2k_BT/\hbar \omega _p)\lambda ,\quad \lambda >2Ze^2/\hbar v_T.
\end{array}
\right.
\end{equation}

As it follows from the expression (29), for low-intensity magnetic fields ($%
\beta <1$), the ELR tends to the well-known Bohr result [24]

\begin{equation}
S_B=\frac{Z^2e^2\omega _p^2}{u_0}\ln \left( \frac{k_{\max }u_0}{\omega _p}%
\right) .
\end{equation}

Meanwhile, for the high-intensity magnetic fields ($\beta >1$), the
expression (29) tends to a constant value $q^2\omega _p^2/2u_0$, which also
follows from Eq. (14) when thermal motion of electrons is ignored. For
arbitrary values of $\beta $, the ELR do not exceed the Bohr losses (see
Fig. 2).

\subsection{Transversal motion of a particle ($\vartheta =\pi /2$)}

In the case of the transversal motion of a particle, $u_0=0$, and the
general expression (8) becomes

\begin{equation}
S=\frac{2Z^2e^2\Omega _c^2}{\pi v}\sum_{n=1}^\infty nQ_n(s)\limfunc{Im}
\left[ \frac{-1}{\varepsilon (n\Omega _c)T(n\Omega _c)}\right] ,
\end{equation}
where $s=k_{\max }a$,

\begin{equation}
T(\omega )=\sqrt{\frac{\left| P(\omega )\right| +{\rm Re}P(\omega )}2}+i{\rm %
sgn}\left[ {\rm Im}P(\omega )\right] \sqrt{\frac{\left| P(\omega )\right| -%
{\rm Re}P(\omega )}2},
\end{equation}
\begin{equation}
Q_\nu (s)=\pi \int_0^sdxJ_\nu ^2(x).
\end{equation}
Function $Q_\nu (s)$ is examined in the Appendix, where asymptotic values
are also given. The function $Q_\nu (s)$ is shown to be exponentially small
at $\nu >s$. Therefore, the series entering Eq. (32) is cut at $n_{\max
}\simeq s$ and the ELR is determined by harmonics having $n<n_{\max }$.

Let us study Eq. (32) in the range of strong magnetic fields. Two cases must
be mentioned here.

(i) $c=\omega _c/\Omega _c$ is a fraction. In this case, from Eq. (32) we
find

\begin{equation}
S\simeq \frac{Z^2e^2\omega _p^2}{\pi v}\frac \nu {\Omega
_c}\sum_{n=1}^\infty \frac 1{n^2}Q_n(s)\left[ 1+\frac{n^4}{\left(
n^2-c^2\right) ^2}\right] .
\end{equation}
From Eq. (35) it follows that the energy loss decreases inversely
proportional to the magnetic field.

(ii) $c=1$ (electron test particle). From Eq. (32) in this case we find

\begin{equation}
S\simeq \frac{Z^2e^2\omega _p^2}{\pi v}\frac{\Omega _c}\nu Q_1(s).
\end{equation}
In this case the ELR increases proportionally to the magnetic field.

The above examples of the asymptotic ELR dependence on the value of the
magnetic field show strong dependence of ELR on the mass of a test particle
in the case when the magnetic field is sufficiently strong.

From Eq. (32) it is easy to trace qualitatively the behavior of energy
losses as a function of magnetic field in the general case. Thus, as it
follows from Eq. (32), the ELR is maximal for those values of the magnetic
field for which $\varepsilon (n\Omega _c)$ is small. The smallness $%
\varepsilon (n\Omega _c)$ means that the dependence of the ELR from the
magnetic field reveals maxima at integer values of parameter $b=a/\lambda
_p\equiv \omega _p/\Omega _c$, where $\lambda _p=2\pi v/\omega _p$ is the
plasma oscillations' wavelength.

Figure 3 shows ELR to Bohr ELR ratio as a function of parameter $b$ in two
cases: for proton (dotted line) and electron (solid line) test particle. The
plasma and/or particle parameters are taken equal to $T=100\ {\rm eV}$, $%
n_0=10^{18}\ {\rm cm}^{-3}$, $\upsilon /\omega _{pe}=0.01$, and $\lambda =10$%
. As it follows from Fig. 3, ELR oscillates as a function of magnetic field
and many times exceeds the usual Bohr ELR.

\section{SUMMARY}

The purpose of this work was to analyze the energy loss rate (ELR) of a
charged particle in a magnetized classical plasma. Larmor rotation of a test
particle in a magnetic field was taken into account. A general expression
obtained for ELR was analyzed in three particular cases: in a Maxwellian
plasma under a strong magnetic field; in a cold plasma when the particle
moves along the magnetic field; and in a cold plasma when the particle moves
across the magnetic field.

The energy loss in a Maxwellian plasma, both in the presence of a strong
magnetic field and in its absence, is conditioned by the induced plasma
waves. In the presence of a strong magnetic field, the dispersion of plasma
oscillations is perceptibly altered. From the expression (10) one may see
that the frequency and the damping rate of these waves depend on the
direction of spreading relative to the magnetic field. The maximal frequency
of these waves is reached when they are spread along the magnetic field.
Across the magnetic field, they cannot be spread. It can be noticed that for
the electron plasma oscillations, these effects are analyzed in detail in
Refs. [17,27,28].

From the results obtained in Sec. IV, one may conclude that the ELR
essentially depends on the particle's incident angle with respect to
magnetic field. In the case of longitudinal motion ($\vartheta =0$), the ELR
is less than or comparable with Bohr's result, and in the limit of strong
magnetic fields, ELR depends only on the density of the plasma. When the
particle moves across the magnetic field ($\vartheta =\pi /2$), the latter
essentially affects the ELR value. First, ELR has an oscillatory character
of dependence on a magnetic field, becoming maximal at integer values of
parameter $b=\omega _p/\Omega _c$ (the ratio of Larmor circle length and
plasma wave wavelength). Second, ELR in the magnetized plasma at $\vartheta
=\pi /2$ is much greater than the Bohr result. Third, the strong dependence
of ELR on the mass of the test particle can be seen when the magnetic field
is sufficiently strong. If thermal motion of plasma electrons is considered,
the results obtained in Sec. IV will be preserved in general. However, the
new effects related to the increased number of normal plasma modes will
originate. In particular, at $\vartheta =\pi /2$, the new mechanism of
stopping could be expected, namely stopping by excitation of the Bernstein
oscillations [21].

\begin{center}
{\bf ACKNOWLEDGMENT}
\end{center}

The author would like to thank Professor Claude Deutsch for valuable help
and discussions.

\begin{center}
{\bf APPENDIX}
\end{center}

Let us examine the properties of function $Q_\nu (s)$ determined by Eq.
(34). To find the asymptotic value of that function at $s\gg 1$ and $s>\nu $%
, we partition the area of integration in Eq. (34) into areas $x<\nu $ and $%
\nu <x<s$ and use the asymptotic presentation of the Bessel function at $%
x>\nu $ [29]. Thus, we find

\begin{equation}
Q_\nu (s)\simeq q_\nu +\ln \frac s\nu +\cos (\pi \nu )\left[ {\rm si}(2s)-%
{\rm si}(2\nu )\right] -\sin (\pi \nu )\left[ {\rm ci}(2s)-{\rm ci}(2\nu
)\right] ,  \tag{A1}
\end{equation}
where ${\rm si}(z)$ and ${\rm ci}(z)$ are integral sine and cosine,
respectively,

\begin{equation}
q_\nu =\pi \int_0^\nu dxJ_\nu ^2(x).  \tag{A2}
\end{equation}
Numbers $q_\nu $ are less than 1, and slowly fall off as the $\nu $
increases. Here we point out some values of $q_\nu $: $q_1\simeq 0.225,$ $%
q_{20}\simeq 0.096,$ $q_{100}\simeq 0.057$.

At $s<\nu $, the argument of the Bessel function is lower than the index. In
this case, the Bessel function is exponentially small, and at a fixed value
of $s$, $Q_\nu (s)$ exponentially vanishes as $\nu $ increases.

\begin{enumerate}
\item[{[1]}]  J. Lindhard, Mat. Fys. Medd. K. Dan. Vidensk. Selsk. {\bf 28,}
8 (1954).

\item[{[2]}]  J. Neufeld and R. H. Ritchie, Phys. Rev. {\bf 98,} 1632 (1955).

\item[{[3]}]  F. Perkins, Phys. Fluids {\bf 8,} 1361 (1965).

\item[{[4]}]  M. M. Basko, Fiz. Plazmy {\bf 10,} 1195 (1984) [Sov. J. Plasma
Phys. {\bf 10,} 689 (1984)].

\item[{[5]}]  Th. Peter and J. Meyer-ter-Vehn, Phys. Rev. A {\bf 43,} 1998
(1991).

\item[{[6]}]  I. M. Bespalov, A. V. Bashman, S. L. Leshkevich, A. Ya.
Polishchuk, A. Yu. Seval'nikov, and V. E. Fortov, Fiz. Plazmy {\bf 17,} 205
(1991) [Sov. J. Plasma Phys. {\bf 17,} 199 (1991)].

\item[{[7]}]  J. D' Avanzo, M. Lontano, and P. F. Bortignon, Phys. Rev. E 
{\bf 47,} 3574 (1993).

\item[{[8]}]  C. Couillaud, R. Deicas, Ph. Nardin, M. A. Beuve, J. M.
Guihaum\'e, M. Renaud, M. Cukier, C. Deutsch, and G. Maynard, Phys. Rev. E 
{\bf 49,} 1545 (1994).

\item[{[9]}]  V. P. Silin, {\it Introduction to the Kinetic Theory of Gases}
(Nauka, Moscow, 1971), Chap. 10 (in Russian).

\item[{[10]}]  I. A. Akhiezer, Zh. Eksp. Teor. Fiz. {\bf 40}, 954 (1961)
[Sov. Phys. JETP {\bf 13}, 667 (1961)].

\item[{[11]}]  N. Honda, O. Aona, and T. Kihara, J. Phys. Soc. Jpn. {\bf 18}%
, 256 (1963).

\item[{[12]}]  R. M. May and N. F. Cramer, Phys. Fluids {\bf 13}, 1766
(1970).

\item[{[13]}]  G. G. Pavlov and D. G. Yakovlev, Zh. Eksp. Teor. Fiz. {\bf 70}%
, 753 (1976) [Sov. Phys. JETP {\bf 43}, 389 (1976)].

\item[{[14]}]  J. G. Kirk and D. J. Galloway, Plasma Phys. {\bf 24}, 339
(1982).

\item[{[15]}]  S. V. Bozhokin and \'E. A. Choban, Fiz. Plazmy {\bf 10}, 779
(1984) [Sov. J. Plasma Phys. {\bf 10}, 452 (1984)].

\item[{[16]}]  E. M. Lifshitz and L. P. Pitaevski\'\i , {\it Physical
Kinetics} (Nauka, Moscow, 1979) (in Russian).

\item[{[17]}]  N. Rostoker and M. N. Rosenbluth, Phys. Fluids {\bf 3,} 1
(1960).

\item[{[18]}]  A. A. Ware and J. C. Wiley, Phys. Fluids B {\bf 5,} 2764
(1993).

\item[{[19]}]  S. T. Butler and M. J. Buckingham, Phys. Rev. {\bf 126}, 1
(1962).

\item[{[20]}]  Yu. V. Gott, {\it Interaction of Particles with Matter in
Plasma Research} (Atomizdat, Moscow, 1978), Chap. 2, Sec. 7 (in Russian).

\item[{[21]}]  I. B. Bernstein, Phys. Rev. {\bf 109,} 10 (1958).

\item[{[22]}]  T. Hagfors, J. Geophys. Res. {\bf 66}, 1699 (1961).

\item[{[23]}]  D. B. Fried and S. D. Conte, {\it The Plasma Dispersion
Function }(Academic Press, New York, 1961).

\item[{[24]}]  N. Bohr, Philos. Mag. {\bf 30}, 581 (1915).

\item[{[25]}]  N. A. Krall and A. W. Trivelpiece, {\it Principles of Plasma
Physics} (McGraw-Hill, New York, 1973).

\item[{[26]}]  A. F. Alexandrov, L. S. Bogdankevich, and A. A. Rukhadze, 
{\it Principles of Plasma Electrodynamics} (Springer-Verlag, New York, 1984).

\item[{[27]}]  N. D. Mermin and E. Canel, Ann. Phys. (N.Y.) {\bf 26,} 247
(1964).

\item[{[28]}]  V. Celli and N. D. Mermin, Ann. Phys. (N.Y.) {\bf 30,} 249
(1964).

\item[{[29]}]  I. S. Gradshteyn and I. M. Ryzhik, {\it Table of Integrals,
Series and Products} (Academic, New York, 1980).
\end{enumerate}

\begin{center}
\newpage\ {\bf Figure Captions}
\end{center}

Fig.1. ELR (in MeV/sec) of a proton as a function of the dimensionless
parameter $\lambda =u_0/v_T$ in the case when the particle moves in
Maxwellian plasma ($T=10\ {\rm eV}$, $n_0=10^{14}\ {\rm cm}^{-3}$) placed in
a strong magnetic field for two values of $B_0$: $B_0=50\ {\rm kG}$ (dotted
line) and $B_0=80\ {\rm kG}$ (solid line).

Fig.2. Dependence of function $R=S/S_B$ on the dimensionless magnetic field $%
\beta =\omega _c/\omega _p$ in the case when the particle moves along the
magnetic field for the values of parameter $\lambda =5$ (dotted line) and $%
\lambda =10$ (solid line). Plasma parameters are taken equal to $T=100\ {\rm %
eV}$ and $n_0=10^{22}\ {\rm cm}^{-3}$, while $Z=1$ for the test particle.

Fig.3. Dependence of a function $R=S/S_B$ on the dimensionless parameter $%
b=\omega _p/\Omega _c$ for proton (dotted line) and electron (solid line).
Parameters are taken equal to $T=100\ {\rm eV}$, $n_0=10^{18}\ {\rm cm}^{-3}$%
, $\lambda =10$, and $\nu /\omega _p=0.01$.

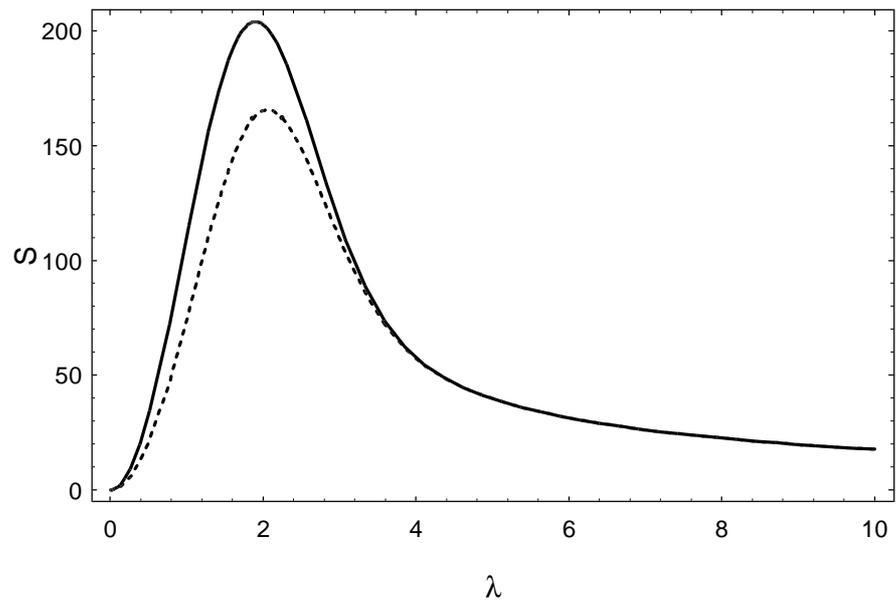
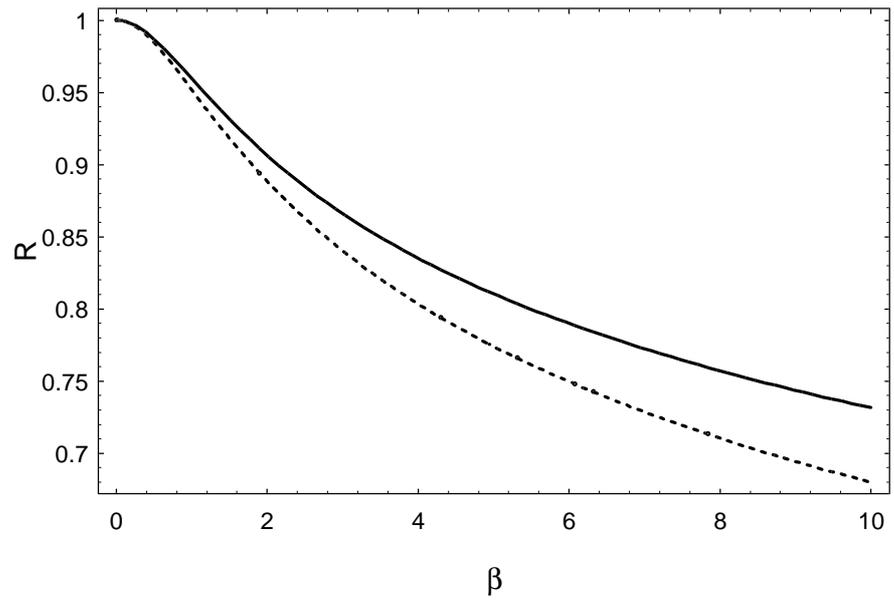
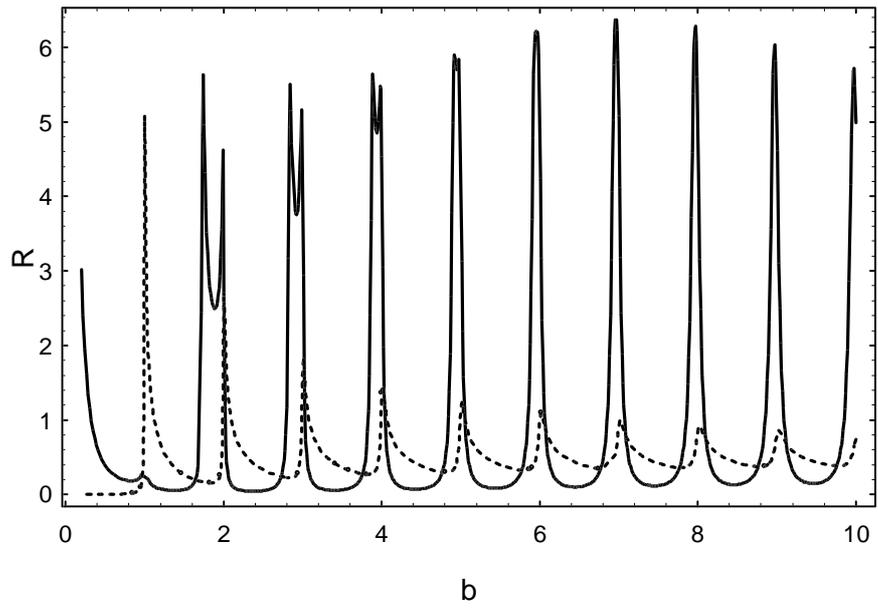
\end{document}